\documentclass[copyright,creativecommons]{eptcs}
\usepackage{graphicx}
\usepackage{amssymb}

\providecommand{\tabularnewline}{\\}

\providecommand{\event}{}
\providecommand{\volume}{??}
\providecommand{\anno}{20??}
\providecommand{\firstpage}{1}
\usepackage{stmaryrd}
\usepackage{breakurl}

\newcommand{\LUSTRE}{{\tt LUSTRE}}

\newcommand{\Esterel}{{\tt Esterel}}

\newcommand{\Bluespec}{{\tt Bluespec}}

\newcommand{\A}{{\tt A}}
\newcommand{\B}{{\tt B}}
\newcommand{\C}{{\tt C}}
\newcommand{\switch}{{\tt S}}

\newcommand{\LA}{{\tt L_A}}
\newcommand{\LB}{{\tt L_B}}

\newcommand{\iOne}{{\tt i0}}
\newcommand{\iTwo}{{\tt i1}}
\newcommand{\oOne}{{\tt o0}}
\newcommand{\oTwo}{{\tt o1}}

\newcommand{\PhOne}{{\tt Ph1}}
\newcommand{\PhTwo}{{\tt Ph2}}

\newcommand{\FIFO}{{\tt FIFO}}

\newcommand{\ChaLo}{{\tt ChaLo}}

\newcommand{\ChaChaLo}{{\tt Cha^{2}Lo}}
\newcommand{\ChaFourLo}{{\tt Cha^{4}Lo}}
\newcommand{\ChaEightLo}{{\tt Cha^{8}Lo}}
\newcommand{\ChaNLo}{{\tt Cha^{n}Lo}}
\newcommand{\ChaNHLo}{{\tt Cha^{n}_{h}Lo}}
\newcommand{\N}{{\tt n}}
\newcommand{\h}{{\tt h}}
\newcommand{\ChaSpeLo}{{\tt Cha^{4}_{3}Lo}}

\newcommand{\TLAplus}{{\tt TLA^{+}}}

\title{Expressing the Behavior of Three Very Different Concurrent Systems by Using Natural Extensions of Separation Logic}

\author{Edgar G. Daylight
\institute{a.k.a. K. Van Oudheusden,\\ Institute of Logic, Language, and Computation,\\ University of Amsterdam, The Netherlands}
\email{egdaylight@yahoo.com}
\and
Sandeep K. Shukla
\institute{Department of Electrical\\ \& Computer Engineering,\\ Virginia Tech., USA}
\email{shukla@vt.edu}
\and
Davide Sergio
\institute{Institute of Logic, Language, and Computation,\\ University of Amsterdam, The Netherlands}
\email{D.Sergio@student.uva.nl}
}
\def\titlerunning{Using Natural Extensions of Separation Logic}
\def\authorrunning{E.G. Daylight \& S.K. Shukla \& D. Sergio}
\begin{document}
\maketitle
\begin{abstract}
Separation Logic is a non-classical logic used to verify pointer-intensive
code. In this paper, however, we show that Separation Logic, along
with its natural extensions, can also be used as a specification language
for concurrent-system design. To do so, we express the behavior of
three very different concurrent systems: a Subway, a Stopwatch, and
a $2\!\times\!2$ Switch. The Subway is originally implemented in
$\LUSTRE$, the Stopwatch in $\Esterel$, and the $2\!\times\!2$
Switch in $\Bluespec$. 
\end{abstract}

\section{\label{sec:Introductionkvo}Introduction }

Concurrent systems, specified today, can have very different properties.
Depending on these properties, a practical specification language
is chosen. For  instance, consider a designer who can choose between
the synchronous language $\Esterel$ and the guarded-command language
$\Bluespec$ in order to specify the \emph{modal} behavior of a Stopwatch,
on the one hand, and the \emph{shared-memory} behavior of a $2\!\times\!2$
Switch, on the other hand. The designer will typically choose $\Esterel$
for the Stopwatch and $\Bluespec$ for the $2\!\times\!2$ Switch
and not the other way around. In other words, while it is of course
theoretically possible to express modal behavior with $\Bluespec$
and shared-memory behavior with $\Esterel$, it is --in terms of practical
expressiveness-- not interesting to do so.

The statements in the previous paragraph are based on {}``common
design experience'', not on a formal metric of practical expressiveness.
To the best of our knowledge, such a metric is not available in the
current literature%
\footnote{Note that the conciseness of a specification is too simplistic a metric:
a lengthy specification $S_{0}$ can be preferred over a short specification
$S_{1}$ if, for instance, $S_{0}$ explicitly captures a design requirement
that is only implicitly present in $S_{1}$.%
}. In this paper we do not try to find such a metric either, for we
believe it is wiser to first obtain many specifications of various
systems using different specification languages and to compare them
based on intuitive notions of {}``practical expressiveness''. Based
on these informal comparisons, we can then search for a metric that
is both well defined and practically relevant. 

In this paper we choose the formalism of Separation Logic and its
natural extensions to express the behavior of three very different
systems:

\begin{itemize}
\item A Subway system, originally specified with $\LUSTRE$~\cite{HalbwachsSubway}.
\item A Stopwatch, originally specified with $\Esterel$~\cite{HalbwachsBook}.
\item A $2\!\times\!2$ Switch, originally specified with $\Bluespec$~\cite{BSwhitepaperSwitch}.
\end{itemize}
Our specifications are based on an analogy that we make with photography,
explained below. The analogy is formalized by means of Separation
Logic~\cite{O'Hearn,Reynolds}. This logic, in turn, is an extension
of Hoare Logic and is typically not used in the way we use it in this
paper, i.e. as a specification language.

\subsection*{An Analogy with Photography}

\begin{figure}[t]
\begin{center}\includegraphics[%
  scale=0.5]{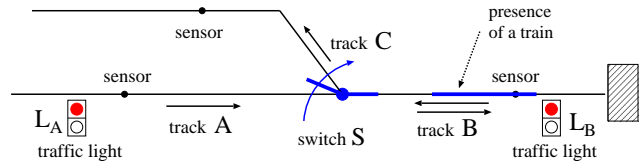}\end{center}

\caption{\label{fig:SubwayScenario}A scenario of the Subway system}
\end{figure}

Given a concurrent system such as the Subway system in Figure~\ref{fig:SubwayScenario},
we make the following analogy with photography. Let various photographers
be assigned to different locations in Figure~\ref{fig:SubwayScenario}.
By taking consecutive camera snapshots, each photographer captures
local change of some part of the Subway. Then, by combining all local
changes, we obtain a complete specification of the Subway. 

For instance, suppose photographer $\PhOne$ is assigned to take snapshots
of track~$\A$ in Figure~\ref{fig:SubwayScenario} while photographer
$\PhTwo$ is assigned to track~$\B$. $\PhOne$ can, by taking one
snapshot, either observe the presence of a train on track~$\A$,
denoted by $1@A$, or the vacancy of track~$\A$, denoted by $0@A$.
By taking two consecutive snapshots, $\PhOne$ can observe four possible
changes: $(a@A,\, b@A)$ with $a,b\in\{0,1\}$ --where we shall use
$(a,\, b)@A$ to abbreviate $(a@A,\, b@A)$. For example, $\PhOne$
may observe the arrival of a train on $\A$, denoted by $(0,1)@A$.
Likewise, $\PhTwo$ may observe the continuous vacancy of track~$\B$,
denoted by $(0,0)@B$. By combining the observations of $\PhOne$
and $\PhTwo$, we obtain the composite change $(0,1)@A$ $*$ $(0,0)@B$,
describing a system in which a train arrives on $\A$ while, simultaneously,
track~$\B$ is vacant. 

The example, presented above, can be extended by adding more photographers,
as we shall illustrate in Section~\ref{sec:Subway} when discussing
the Subway in more detail. In addition, we can generalize the notions
of `snapshot' and `change' to the notion of `change of change'. This
extension will be needed when specifying the modal behavior of a Stopwatch
in Section~\ref{sec:Stopwatch}. In terms of the analogy, the photographer
capturing a scene by means of `change', has become a camera man, capturing
the change from one scene to another. Another generalization is needed
when specifying the $2\!\times\!2$ Switch in Section~\ref{sec:Switch}.
There, the concept of `snapshot' is generalized to that of an `hierarchical
snapshot', implying that each photographer can zoom in on specific
details of the concurrent system under investigation. Consequently,
hierarchical change is used (instead of plain change) to capture the
concurrent behavior of the Switch. 

\subsection*{Related Work}

Our analogy with photography is formalized in this paper by using
the following embarrassingly simple logics. First, the Logic of Snapshots
is merely an instance of Separation Logic's assertion language using
the $@$ primitive of Ahmed et al.~\cite{Walker1} instead of the
usual points-to predicate~\cite{O'Hearn,Reynolds}. The key point
is that formulae denote unary predicates over snapshots ($shot$)
of the system state. The second logic is $\ChaLo$, the Logic of Change.
It is basically Yang's {}``Relational Separation Logic''~\cite{Yang07RelationalSepLog}
where formulae denote binary relations $cha$ of the form $\left(shot_{in},shot_{out}\right)$
rather than unary predicates. For notational convenience we let $\left(f\! st\: cha\right)$
refer to $shot_{in}$ and $\left(snd\: cha\right)$ to $shot_{out}$.
The third logic is $\ChaChaLo$ where formulae denote relations on
relations over snapshots, i.e. sets of elements of the form: $\left(cha_{in},cha_{out}\right)$.
But the semantics will require that $\left(snd\: cha_{in}\right)$
and $\left(f\! st\: cha_{out}\right)$ are always equal (or else completely
irrelevant), so formulae actually denote triples of snapshots. Hence,
$\ChaChaLo$ is a straightforward adaptation of $\ChaLo$ from binary
to ternary relations. Continuing in the same manner, we then present
$\ChaFourLo$, denoting elements of the form: $\left(\left(cha_{1},cha_{2}\right),\left(cha_{3},cha_{4}\right)\right)$
where $\left(snd\: cha_{2}\right)$ is equal to $\left(f\! st\: cha_{3}\right)$.
Further extrapolation results in $\ChaEightLo$ and, in general, in
$\ChaNLo$ with $\N$ $=$ $2^{k}$ and $k$ a strictly positive integer.
In summary, it is more the use of the logical definitions that is
new and interesting, rather than the definitions themselves.

The formalism in this paper abides by the Synchronous Hypothesis~\cite{Berry-Gonthier-ScCompProg92,SynchHypothesis}.
To illustrate this, consider $\left(0,1\right)@A\:\ast\:\left(G,R\right)@L_{A}$,
which can be read operationally as follows: {}``When track~$\A$'s
sensor senses the arrival of a train, the Subway system responds by
turning traffic light~$\LA$ from green ($G$) to red ($R$).''
Operationally, it makes no sense to reason in the opposite direction;
i.e. by starting with the light and concluding with $\A$'s sensor.
Thus, we have an ordering from $\left(0,1\right)@A$ to $\left(G,R\right)@L_{A}$.
But, since $\ast$ requires that both occur simultaneously, $\left(G,R\right)@L_{A}$
has to be an \emph{instantaneous} reaction to $\left(0,1\right)@A$.

The analogy with photography, resulting in the concepts of `snapshot',
`change', and `change of change', sets it apart from other well-established
formalisms, such as Statecharts~\cite{Statecharts}, Communicating
Sequential Processes~\cite{HoareComSeqProc}, the $\pi$-calculus~\cite{MilnerPicalculus},
spatial logics (e.g. \cite{CairesCardelli}), and process algebras~\cite{FokkinkBook},
just to name a few. Lack of space prevents us from delving into these
other formalisms here.

\section{\label{sec:Subway}Subway}

We introduce the Logic of Snapshots and its extension $\ChaLo$ (i.e.
the Logic of Change) to specify a Subway. The Logic of Snapshots is
system dependent. That is, we shall introduce syntax for snapshots
that depends on the Subway. Later, when discussing the Stopwatch (Section~\ref{sec:Stopwatch})
and the Switch (Section~\ref{sec:Switch}), we shall introduce other
syntax. $\ChaLo$, on the other hand, is only defined once. 

This section consists of three parts. Section~\ref{sub:Design-Intent-Subway}
presents the design intent of the Subway. Section~\ref{sub:Some-Illustrations-of-ChaLo-Subway}
illustrates how the Logic of Snapshots and $\ChaLo$ can be used to
specify the Subway. Finally, Section~\ref{sub:LogicofChange} presents
the formalization.

\subsection{\label{sub:Design-Intent-Subway}Design Intent}

The objective in Figure~\ref{fig:SubwayScenario} is to design a
Subway system so that a train can enter by track~$\A$, temporarily
use track~$\B$, and then leave by track~$\C$~\cite{HalbwachsSubway}.
At all times, at most one train is present in the Subway system. Seven
state elements constitute the system. Four state elements are inputs:
the sensor values of the tracks~$\A$, $\B$, and $\C$, and the
switch~$\switch$. Three state elements are outputs: the actuator
of the switch~$\switch$ and the two traffic lights $\LA$ and $\LB$.
Each state element is presented below along with its possible values:\\
\begin{tabular}{cllllcccccccccccccccccccc|ccc}
(i)&
&
sensors of $\A$, $\B$, and $\C$&
$\qquad$~~~~&
$0$&
&
$1$&
&
\tabularnewline
(ii)&
&
Sen\_$\switch$, Act\_$\switch$&
&
$AB$&
&
$o\! f\! f$&
&
$BC$\tabularnewline
(iii)&
&
$\LA$, $\LB$&
&
$G$&
&
$R$&
&
\tabularnewline
\end{tabular}{\footnotesize }\\
When a train is on a track (e.g. track~$\B$ in Figure~\ref{fig:SubwayScenario}),
then the corresponding sensor value is $1$, else it is $0$. The
sensor of switch~$\switch$ has the value $AB$ when tracks $\A$
and $\B$ are connected and $BC$ when tracks $\B$ and $\C$ are
connected. The value $o\! f\! f$ occurs when no tracks are connected,
as is the case in Figure~\ref{fig:SubwayScenario}. The actuator
of switch~$\switch$ has the value $AB$ when the switch is being
steered in order to (eventually) connect tracks $\A$ and $\B$. Similarly,
the value is $BC$ when connecting tracks $\B$ and $\C$, as illustrated
in Figure~\ref{fig:SubwayScenario} by the arrowed arc. The actuator
has the value $o\! f\! f$ when the switch is not being steered (i.e.
typically when two tracks are connected). Traffic lights can either
be green ($G$) or red ($R$). Green light~$\LA$ allows a train
to enter track~$\A$ from the left. Green light~$\LB$ allows a
train to depart from track~$\B$ by moving backwards (preferably
onto track~$\C$!).

\subsection{\label{sub:Some-Illustrations-of-ChaLo-Subway}Some Specifications}

We now illustrate the Logic of Snapshots and $\ChaLo$ by presenting
some specifications of the Subway. Since the complete formalization
of the Logic of Snapshots is straightforward but lengthy, we merely
illustrate it below.  The more important logic $\ChaLo$, on the
other hand, is illustrated below and formally defined in Section~\ref{sub:LogicofChange}. 

As a first example, consider the following snapshot specification
in the Logic of Snapshots:\\
\begin{tabular}{ccl}
(1)&
~~~&
$0@A\:*\:1@B\:*G@L_{B}\:*\: BC@SenS$\tabularnewline
\end{tabular}\\
It partially describes a particular instance of the Subway: track~$\A$
is vacant, track~$\B$ is occupied, traffic light~$\LB$ is green
($G$) and hence granting exit to the train on track~$\B$. The switch~$\switch$
is, based on its sensor ($SenS$), connecting track~$\B$ with track~$\C$.
The snapshot is partial because it does not capture the status of
track~$\C$, the traffic light~$\LA$, and the actuator of the switch~$\switch$.

Expression~(1) is a syntactic abbreviation for:\\
\begin{tabular}{ccl}
(2)&
~~~&
$<0@A*1@B,\: BC@SenS,\: emp,\: G@L_{B}>$\tabularnewline
\end{tabular}\\
which is a tuple of four snapshot expressions. The first entry $0@A*1@B$
describes the states of the tracks, the second describes the switch's
sensor, the third describes the switch's actuator, where $emp$ abbreviates
{}``empty'', and the fourth describes the traffic lights. The meaning
of (2) is described next.

Let $T\! r$ $=$ $\{ f\! alse,\, true\}$ denote the set of truth
values, $V\! ar$ a set of variables, and $V\! al$ a set of values:
$V\! al$ $=$ $\left\{ 0,\,1\right\} \,\cup\left\{ AB,\, BC,\, o\! f\! f\right\} \,\cup\,\left\{ R,\, G\right\} $
and $V\! al_{\bot}=V\! al\cup\left\{ \bot\right\} $. The set of assignment
functions is $Asgmt$ $:=$ $V\! ar\rightarrow V\! al_{\bot}$. Let
$s$ denote an assignment function, i.e. $s\in Asgmt$. Then, the
semantical interpretation of (2), using $s$, results in a semantic
snapshot $(shot_{1},\, shot_{2},\, shot_{3},\, shot_{4})$: \\
\begin{tabular}{llllll}
(3)&
~~~&
\multicolumn{4}{l}{$s,\,(shot_{1},\, shot_{2},\, shot_{3},\, shot_{4})$}\tabularnewline
(4)&
&
&
\multicolumn{3}{l}{$\models\quad<0@A*1@B,\: BC@SenS,\: emp,\: G@L_{B}>$}\tabularnewline
(5)&
&
iff&
&
$s,\, shot_{1}\;\models\;0@A*1@B$&
and\tabularnewline
(6)&
&
&
&
$s,\, shot_{2}\;\models\; BC@SenS$&
and\tabularnewline
(7)&
&
&
&
$s,\, shot_{3}\;\models\; emp$&
and\tabularnewline
(8)&
&
&
&
$s,\, shot_{4}\;\models\; G@L_{B}$&
\tabularnewline
\end{tabular}\\
That is, each local semantic snapshot $shot_{i}$ with $i\in\{1,2,3,4\}$
models the corresponding syntactic snapshot as a function. I.e., $shot_{1}$
is a function that maps $A$ to $0$, $B$ to $1$, and $C$ to $\bot$.
Function $shot_{2}$ maps $SenS$ to $BC$. Function $shot_{3}$ maps
$ActS$ to $\bot$, since no information ({}``empty'' $emp$) is
present about the actuator $ActS$. Function $shot_{4}$ maps $L_{A}$
to $\bot$ and $L_{B}$ to $G$. Finally, we remark that $(shot_{1},\, shot_{2},\, shot_{3},\, shot_{4})$
$\in$ $S\! nshot$ where $S\! nshot$ is the domain of semantic snapshots.

As a second example, we illustrate the difference between Separation
Logic's spatial conjunction~$*$ and classical conjunction~$\wedge$.
Let us take (5) and replace $*$ by $\wedge$, then we have:\\
\begin{tabular}{cllllll}
(9)&
~~~&
\multicolumn{5}{l}{$s,\, shot_{1}\;\models\;0@A\wedge1@B$}\tabularnewline
(10)&
&
iff&
&
$s,\, shot_{1}\;\models\;0@A$&
and&
\tabularnewline
(11)&
&
&
&
$s,\, shot_{1}\;\models\;1@B$&
&
\tabularnewline
\end{tabular}\\
Now, (10) states that $shot_{1}$ is a function mapping $A$ to $0$
and $B$ and $C$ to $\bot$. On the other hand, (11) states that
$shot_{1}$ maps $B$ to $1$ and $A$ and $C$ to $\bot$. This is
clearly not possible, so $\wedge$ is used incorrectly in (9). 

The previous example shows that $\wedge$ can not replace $*$ without
altering the intended meaning. This is due to the chosen semantics
of $@$:  $0@A$ \emph{only} describes the state of track~$\A$.
In the alternative classical semantics, $0@A$ would describe the
\emph{complete} state of all three tracks in the Subway system, with
the additional knowledge that track~$\A$ is vacant%
\footnote{It is of course possible to avoid the use of $*$ in this paper by
redefining the meaning of $@$ in accordance to the classical semantics,
but the purpose of this paper is to use Separation Logic for case
studies, such as the Subway system, for which it was not initially
intended. %
}. The same arguments also hold for change, such as $\left(0,1\right)@A$. 

For a third example, recall the photographers $\PhOne$ and $\PhTwo$
from Section~\ref{sec:Introductionkvo}. When both photographers
combine their observations, they conclude that, in accordance to a
correctly-behaving Subway system, the following implication has to
hold: $\left(0,1\right)@A$ $\rightarrow$ $\left(0,0\right)@B$.
In words: if a train arrives on track~$\A$, then, at the same time,
track~$\B$ remains vacant. That is, it is impossible for $\PhOne$
to observe $\left(0,1\right)@A$ while $\PhTwo$ observes, say, $\left(1,0\right)@B$.

The implication $\left(0,1\right)@A$ $\rightarrow$ $\left(0,0\right)@B$
is an abbreviation for: \\
$\left(0,1\right)@A\:*\:\exists x\exists y\,(x,y)@B$ ~~$\Rightarrow$~~
$\exists x'\exists y'\,(x',y')@A\:*\:\left(0,0\right)@B$, \\
where we have ensured that the same state elements ($A$ and $B$)
are present on the left- and righthand side of $\Rightarrow$. That
is, $\rightarrow$ is defined here (by example) in terms of $\Rightarrow$,
which, in turn, is defined formally in the next section (cf. Table~\ref{cap:SyntaxSemantics2Upper}). 

As a fourth and final example, consider $\left(0,1\right)@A\leftrightarrow\left(G,R\right)@L_{A}$,
which is an abbreviation for $\left[\left(0,1\right)@A\rightarrow\left(G,R\right)@L_{A}\right]$
$\wedge$ $\left[\left(G,R\right)@L_{A}\rightarrow\left(0,1\right)@A\right]$.
It describes the arrival of a train on $\A$ while traffic light $\LA$
turns from green to red. Using this, $\PhOne$, $\PhTwo$, and the
photographer of light $\LA$ can combine ($\otimes$) their observations
as follows: \\
\begin{tabular}{ccl}
(12)&
~~~&
$\left[\:\left(0,1\right)@A\rightarrow\left(0,0\right)@B\:\right]$
$\otimes$ $\left[\:\left(0,1\right)@A\leftrightarrow\left(G,R\right)@L_{A}\:\right]$\tabularnewline
\end{tabular}\\
From this we can, for instance, deduce that $\left(G,R\right)@L_{A}$
implies $\left(0,0\right)@B$.

Similar to $*$ and $\rightarrow$, the use of $\otimes$ aids us
in obtaining a short formal exposition. It could be completely avoided
by only using $*$ and $\wedge$ but at the cost of longer specifications.
It too is formally defined in Table~\ref{cap:SyntaxSemantics2Upper}.

\subsubsection*{Additional Notation }

Let $X_{\bot}:=X\cup\{\bot\}$. For $f\,::\, A_{\bot}\rightarrow B_{\bot}$
we write $f=\underline{\lambda}x.\,\alpha\,$ to denote the mapping:
$f(\bot)=\bot$ and $f(a)=\left[a/x\right]\alpha$ for $a\in A$.
Also, the domain $\left(dom\, f\right)$ of a partial function $f$
is the set of $x$'s such that $f\left(x\right)$ does not equal $\bot$.
In particular, $\left(dom\:\left(\lambda x.\bot\right)\right)$ $=$
$\emptyset$. Finally, for domains $D$ and $E$, let $[D\rightarrow E]$
denote the set $\{ f\mid f::D\rightarrow E\}$. Consider functions
$f,g$ $\in$ $\left[D\rightarrow E_{\bot}\right]$. We define the
operations $\sharp$ and $\centerdot$ as follows:\\
{\footnotesize }\begin{tabular}{lllll}
$\sharp$&
$::$&
\multicolumn{3}{l}{$\left(D\rightarrow E_{\bot}\right)_{\bot}\rightarrow\left(D\rightarrow E_{\bot}\right)_{\bot}\rightarrow T\! r_{\bot}$}\tabularnewline
$\sharp$&
$:=$&
\multicolumn{3}{l}{$\underline{\lambda}f.\,\underline{\lambda}g.\;\left(dom\: f\right)\cap\left(dom\: g\right)==\emptyset$}\tabularnewline
$\centerdot$&
$::$&
\multicolumn{3}{l}{$\left(D\rightarrow E_{\bot}\right)_{\bot}\rightarrow\left(D\rightarrow E_{\bot}\right)_{\bot}\rightarrow\left(D\rightarrow E_{\bot}\right)_{\bot}$}\tabularnewline
$\centerdot$&
$:=$&
\multicolumn{3}{l}{$\underline{\lambda}f.\,\underline{\lambda}g.$ if $f\,\sharp\, g$
then $f\cup g$ else $\bot$}\tabularnewline
\end{tabular}{\footnotesize }\\
For example, if we revisit the partial function $shot_{1}$ in (5).
Then $shot_{1}$ $=$ $shot_{1}^{a}\centerdot shot_{1}^{b}$ where
$shot_{1}^{a}$ is a function that maps $A$ to $0$ and $B$ and
$C$ to $\bot$. Likewise, $shot_{1}^{b}$ maps $B$ to $1$ and $A$
and $C$ to $\bot$. Clearly, $shot_{1}^{a}$ and $shot_{1}^{b}$
have disjoint domains: $shot_{1}^{a}$ $\sharp$ $shot_{1}^{b}$.

\subsection{\label{sub:LogicofChange}Logic of Change}

We are now in a position to present $\ChaLo$, the Logic of Change.
After taking the following four remarks into account, Table~\ref{cap:SyntaxSemantics2Upper}
can be consulted. 

First, we define semantical change as a pair of semantical snapshots:
{\footnotesize }\\
{\footnotesize }\begin{tabular}{rll}
$ch\in Change$&
$:=$&
$S\! nshot\times S\! nshot$\tabularnewline
\end{tabular} {\footnotesize }\\
Second, given semantical changes $ch_{1}$ and $ch_{2}$, the disjointness
($\sharp$) and the combination ($\centerdot$) of $ch_{1}$ and $ch_{2}$
can be defined:\\
{\footnotesize }\begin{tabular}{rclll}
{\footnotesize $ch_{1},ch_{2}$}&
{\footnotesize $\in$}&
\multicolumn{3}{l}{{\footnotesize $Change$}}\tabularnewline
{\footnotesize $ch_{1}$}&
{\footnotesize $=$}&
\multicolumn{3}{l}{{\footnotesize $\left(shot_{1},\, shot'_{1}\right)$}}\tabularnewline
{\footnotesize $ch_{2}$}&
{\footnotesize $=$}&
\multicolumn{3}{l}{{\footnotesize $\left(shot_{2},\, shot'_{2}\right)$}}\tabularnewline
{\footnotesize $ch_{1}\,\sharp\, ch_{2}\;$}&
{\footnotesize iff}&
{\footnotesize $\; shot_{1}\,\sharp\, shot_{2}$}&
{\footnotesize and}&
{\footnotesize $shot'_{1}\,\sharp\, shot'_{2}$}\tabularnewline
{\footnotesize $ch_{1}\centerdot ch_{2}\;$}&
{\footnotesize $:=$}&
\multicolumn{3}{l}{{\footnotesize $\;\left(shot_{1}\centerdot shot_{2},\, shot'_{1}\centerdot shot'_{2}\right)$}}\tabularnewline
\end{tabular}{\footnotesize }\\
Third, the semantics of a $\ChaLo$ formula $P$ is of the form:\\
{\footnotesize }\begin{tabular}{clrllccc|}
\multicolumn{8}{l}{{\footnotesize $s,\,\left(shot_{in},shot_{out}\right)\:\models\: P$
~~~~~or~~~~~ $s,\, ch\:\models\: P$}}\tabularnewline
&
{\footnotesize with }&
{\footnotesize $s$ }&
{\footnotesize $\in$}&
\multicolumn{4}{l}{{\footnotesize $Asgmt$}}\tabularnewline
&
&
{\footnotesize $shot_{in},shot_{out}$}&
{\footnotesize $\in$}&
\multicolumn{4}{l}{{\footnotesize $S\! nshot$}}\tabularnewline
&
&
{\footnotesize $ch$}&
{\footnotesize $=$}&
\multicolumn{4}{l}{{\footnotesize $\left(shot_{in},shot_{out}\right)$}}\tabularnewline
&
&
{\footnotesize $f\! ree\left(P\right)$}&
{\footnotesize $\subseteq$}&
\multicolumn{4}{l}{{\footnotesize $\left(dom\: s\right)$}}\tabularnewline
\end{tabular}{\footnotesize }\\
where $f\! ree\left(P\right)$ denotes the free variables in $P$.
Fourth, an example of an expression relation $Expr\! Rel$ is $x=1$
and its valuation $\left(\left\llbracket Expr\! Rel\right\rrbracket \; s\right)$
amounts to checking whether $\left(s\: x\right)=1$ holds. The trivial
definition of $\left\llbracket Expr\! Rel\right\rrbracket $ is omitted
from this paper.

\begin{table*}

\caption{\label{cap:SyntaxSemantics2Upper}$\ChaLo$}

\begin{center}{\scriptsize }\begin{tabular}{cc}
\hline 
{\scriptsize }\begin{tabular}{clllclcccc|ccc}
{\scriptsize (1)}&
{\scriptsize $P,\: Q$}&
{\scriptsize $::=$}&
\multicolumn{10}{l}{{\scriptsize $Expr\! Rel\:\mid\:\left(S\! nap,\: S\! nap\right)\:\mid\: f\! alse\:\mid\: P;Q\:\mid$}}\tabularnewline
{\scriptsize (2)}&
&
&
\multicolumn{10}{l}{{\scriptsize $P\Rightarrow Q\:\mid\: P\,\ast\, Q\:\mid\: P\,\otimes\, Q\:\mid\:\exists x.\, P$}}\tabularnewline
\multicolumn{13}{l}{{\scriptsize Sugar:}}\tabularnewline
{\scriptsize (3)}&
\multicolumn{12}{c}{{\scriptsize $\left(Expr_{1},Expr_{2}\right)@Place$}}\tabularnewline
&
\multicolumn{12}{c}{{\scriptsize $\equiv$ }}\tabularnewline
&
\multicolumn{12}{c}{{\scriptsize $\left(Expr_{1}@Place,\, Expr_{2}@Place\right)$}}\tabularnewline
{\scriptsize (4)}&
\multicolumn{3}{r}{{\scriptsize $\neg P$}}&
{\scriptsize $\equiv$}&
{\scriptsize $P\Rightarrow f\! alse$}\tabularnewline
{\scriptsize (5)}&
\multicolumn{3}{r}{{\scriptsize $true$}}&
{\scriptsize $\equiv$}&
{\scriptsize $\neg f\! alse$}\tabularnewline
{\scriptsize (6)}&
\multicolumn{3}{r}{{\scriptsize $P\vee Q$}}&
{\scriptsize $\equiv$}&
{\scriptsize $\left(\neg P\right)\Rightarrow Q$}\tabularnewline
{\scriptsize (7)}&
\multicolumn{3}{r}{{\scriptsize $P\wedge Q$}}&
{\scriptsize $\equiv$}&
{\scriptsize $\neg\left(\neg P\,\vee\,\neg Q\right)$}\tabularnewline
{\scriptsize (8)}&
\multicolumn{3}{r}{{\scriptsize $\forall x.\, P$}}&
{\scriptsize $\equiv$}&
{\scriptsize $\neg\exists x.\neg P$}\tabularnewline
{\scriptsize (9)}&
\multicolumn{3}{r}{{\scriptsize $\forall x,y.\, P$}}&
{\scriptsize $\equiv$}&
{\scriptsize $\forall x.\forall y.\, P$}\tabularnewline
\multicolumn{13}{l}{\textbf{\scriptsize Semantics:}}\tabularnewline
\multicolumn{13}{l}{{\scriptsize }\begin{tabular}{lllcccc}
{\scriptsize (10)}&
\multicolumn{6}{l}{{\scriptsize $s,\, ch\:\models\: Expr\! Rel$}}\tabularnewline
&
&
&
{\scriptsize iff}&
&
\multicolumn{2}{c}{{\scriptsize $\left(\left\llbracket Expr\! Rel\right\rrbracket \; s\right)$}}\tabularnewline
\end{tabular}}\tabularnewline
\multicolumn{13}{l}{{\scriptsize }\begin{tabular}{lllcccc}
{\scriptsize (11)}&
\multicolumn{6}{l}{{\scriptsize $s,\,\left(shot_{in},\, shot_{out}\right)\:\models\:\left(S\! nap_{in},\, S\! nap_{out}\right)$}}\tabularnewline
&
&
&
{\scriptsize iff}&
&
&
\tabularnewline
&
\multicolumn{6}{l}{ {\scriptsize $s,\, shot_{in}\:\models\: S\! nap_{in}$ ~~and~~$s,\, shot_{out}\:\models\: S\! nap_{out}$}}\tabularnewline
\end{tabular}}\tabularnewline
\multicolumn{13}{l}{{\scriptsize }\begin{tabular}{llllcccc|}
{\scriptsize (12)}&
{\scriptsize $s,\, ch\:\models\: f\! alse$}&
&
&
\multicolumn{4}{c}{{\scriptsize never}}\tabularnewline
\end{tabular}}\tabularnewline
\multicolumn{13}{l}{{\scriptsize }\begin{tabular}{lllcccc}
{\scriptsize (13)}&
\multicolumn{6}{l}{{\scriptsize $s,\,\left(shot_{in},\, shot_{out}\right)\:\models\: P;Q$}}\tabularnewline
&
&
&
{\scriptsize iff}&
&
&
\tabularnewline
&
\multicolumn{6}{l}{ {\scriptsize $\exists shot_{tmp}\in S\! nshot.$ }}\tabularnewline
&
\multicolumn{3}{l}{{\scriptsize $s,\,\left(shot_{in},\, shot_{tmp}\right)\:\models\: P$}}&
\multicolumn{3}{l}{{\scriptsize and}}\tabularnewline
&
\multicolumn{3}{l}{{\scriptsize $s,\,\left(shot_{tmp},\, shot_{out}\right)\:\models\: Q$}}&
\multicolumn{3}{l}{}\tabularnewline
\end{tabular}}\tabularnewline
\end{tabular}&
{\scriptsize }\begin{tabular}{llllrclllc}
\multicolumn{10}{l}{{\scriptsize }\begin{tabular}{lllcccc}
{\scriptsize (14)}&
\multicolumn{6}{l}{{\scriptsize $s,\, ch\:\models\: P\Rightarrow Q$}}\tabularnewline
&
&
{\scriptsize iff}&
&
&
&
\tabularnewline
&
{\scriptsize if }&
{\scriptsize $s,\, ch\:\models\: P$}&
{\scriptsize then }&
\multicolumn{3}{c}{{\scriptsize $s,\, ch\:\models\: Q$}}\tabularnewline
\end{tabular}}\tabularnewline
\multicolumn{10}{l}{{\scriptsize }\begin{tabular}{lllcccc}
{\scriptsize (15)}&
\multicolumn{6}{l}{{\scriptsize $s,\,\left(shot_{in},\, shot_{out}\right)\:\models\: P\,\ast\, Q$}}\tabularnewline
&
&
&
{\scriptsize iff}&
&
&
\tabularnewline
&
\multicolumn{6}{l}{{\scriptsize $\exists shot_{in}^{1},\, shot_{in}^{2}\in S\! nshot.$}}\tabularnewline
&
&
\multicolumn{4}{l}{{\scriptsize $shot_{in}^{1}\:\sharp\: shot_{in}^{2}$}}&
{\scriptsize and}\tabularnewline
&
&
\multicolumn{4}{l}{{\scriptsize $shot_{in}=shot_{in}^{1}\:\centerdot\: shot_{in}^{2}$}}&
{\scriptsize and}\tabularnewline
&
\multicolumn{6}{l}{{\scriptsize $\exists shot_{out}^{1},\, shot_{out}^{2}\in S\! nshot.$}}\tabularnewline
&
&
\multicolumn{4}{l}{{\scriptsize $shot_{out}^{1}\:\sharp\: shot_{out}^{2}$}}&
{\scriptsize and}\tabularnewline
&
&
\multicolumn{4}{l}{{\scriptsize $shot_{out}=shot_{out}^{1}\:\centerdot\: shot_{out}^{2}$}}&
{\scriptsize and}\tabularnewline
&
\multicolumn{5}{l}{{\scriptsize $s,\,\left(shot_{in}^{1},\, shot_{out}^{1}\right)\:\models\: P$}}&
{\scriptsize and}\tabularnewline
&
\multicolumn{5}{l}{{\scriptsize $s,\,\left(shot_{in}^{2},\, shot_{out}^{2}\right)\:\models\: Q$}}&
\tabularnewline
\end{tabular}}\tabularnewline
\multicolumn{10}{l}{{\scriptsize }\begin{tabular}{lllcccc}
{\scriptsize (16)}&
\multicolumn{6}{l}{{\scriptsize $s,\, ch\:\models\: P\,\otimes\, Q$}}\tabularnewline
&
&
&
{\scriptsize iff}&
&
&
\tabularnewline
&
\multicolumn{6}{l}{{\scriptsize $\exists ch_{0},\, ch_{1},\, ch_{2}\in Change.$}}\tabularnewline
&
&
\multicolumn{4}{l}{{\scriptsize $s,\, ch_{0}\centerdot ch_{1}\models P$}}&
{\scriptsize and}\tabularnewline
&
&
\multicolumn{4}{l}{{\scriptsize $s,\, ch_{0}\centerdot ch_{2}\models Q$}}&
{\scriptsize and}\tabularnewline
&
&
\multicolumn{4}{l}{{\scriptsize $ch_{1}\,\sharp\, ch_{2}$}}&
{\scriptsize and}\tabularnewline
&
&
\multicolumn{4}{l}{{\scriptsize $ch=ch_{0}\centerdot ch_{1}\centerdot ch_{2}$}}&
\tabularnewline
\end{tabular}}\tabularnewline
\multicolumn{10}{l}{{\scriptsize }\begin{tabular}{lllcccc}
{\scriptsize (17)}&
\multicolumn{6}{l}{{\scriptsize $s,\, ch\:\models\:\exists x.\, P$}}\tabularnewline
&
&
&
{\scriptsize iff}&
&
&
\tabularnewline
&
\multicolumn{6}{l}{{\scriptsize $\exists v\in V\! al.\;\; s\left[x\mapsto v\right],\, ch\:\models\: P$}}\tabularnewline
\end{tabular}}\tabularnewline
\end{tabular}\tabularnewline
\hline
\end{tabular}\end{center}
\end{table*}

\section{\label{sec:Stopwatch}Stopwatch}

Our second case study is a Stopwatch, introduced in Section~\ref{sub:Design-Intent-Stopwatch}.
To capture its behavior, we shall introduce the Logic of Change of
Change ($\ChaChaLo$) and similar extensions ($\ChaFourLo$, $\ChaEightLo$,
$\ldots$) in Section~\ref{sub:Logic-of-Change-of-Change}. Finally,
various specifications of the Stopwatch's behavior are presented in
Section~\ref{sub:Some-Specifications-Stopwatch}.

\subsection{\label{sub:Design-Intent-Stopwatch}Design Intent}

The design intent of the Stopwatch is too lengthy to present in plain
English. Therefore, we let our specifications speak for themselves.
They can also be checked against the original $\Esterel$ specification,
presented in~\cite{HalbwachsBook}.  

\begin{figure}
\begin{center}\begin{tabular}{ccccc}
\includegraphics[%
  scale=0.4]{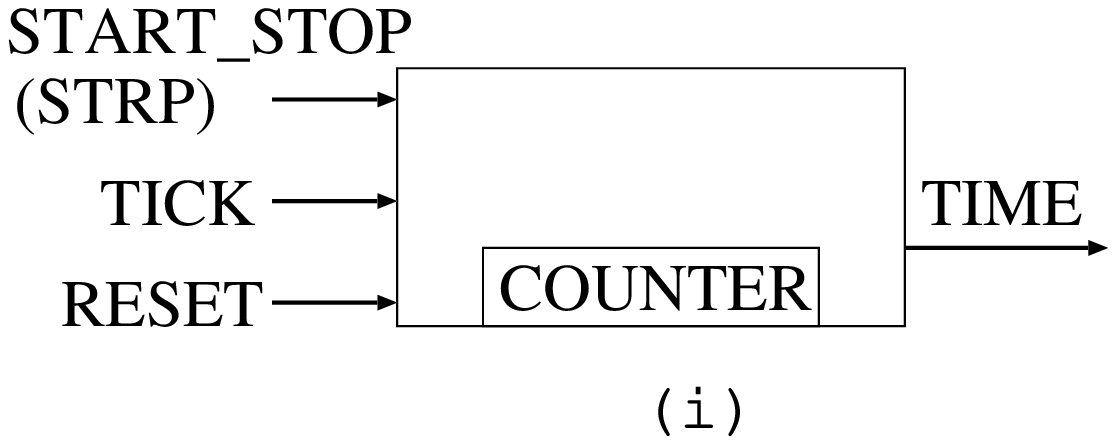}&
~$\quad$~&
\includegraphics[%
  scale=0.4]{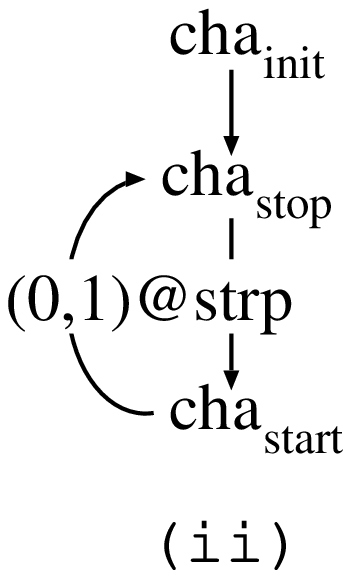}&
~$\quad$~&
\includegraphics[%
  scale=0.4]{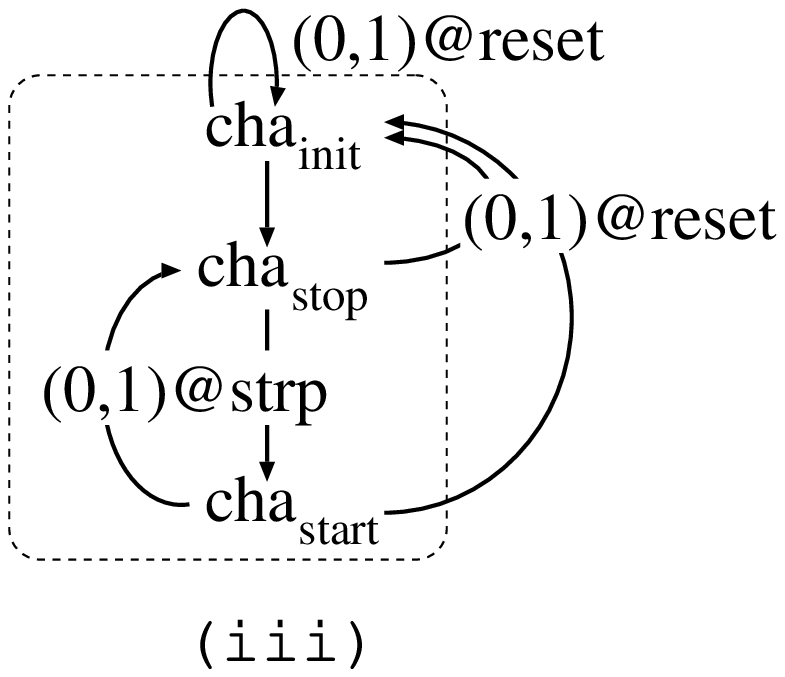}\tabularnewline
\end{tabular}\end{center}

\caption{\label{fig:completeStopwatch}(i)~The Stopwatch, (ii)~the $\ChaChaLo$
diagram for $cha_{basic}^{2}$, and (iii)~the $\ChaFourLo$ diagram
for $cha_{reset}^{4}$.}
\end{figure}

The Stopwatch in Figure~\ref{fig:completeStopwatch}(i) can be briefly
described as follows. The input signals {\tt START\_STOP} (or {\tt STRP}
for short), {\tt TICK}, and {\tt RESET} are \emph{immutable} elements.
That is, their value is completely determined by the external behavior
of the Stopwatch. In fact, {\tt STRP} and {\tt RESET} are buttons
which are pressed ($1$) or depressed ($0$) by the user, and {\tt TICK}
is the signal ($0$ or $1$) of an external clock. The internal register
{\tt COUNTER} and the output signal {\tt TIME} are \emph{mutable}
elements. That is, their value is determined by the internal behavior
of Stopwatch. Finally, we also use an internal register {\tt MODE},
not shown in Figure~\ref{fig:completeStopwatch} to book-keep the
current mode of execution. It too is a mutable element. 

The locations, presented above, can be assigned to the photographers.
We present six examples. First, $\left(0,1\right)@strp$, describing
change from $0@strp$ to $1@strp$, captures the behavior of a user
who presses the {\tt STRP} button. Second, $\left(0,1\right)@strp$
$*$ $\left(0,1\right)@reset$ describes a user who simultaneously
presses both the {\tt STRP} and {\tt RESET} buttons. Third, $(x,x+1)@time$
expresses an increase of {\tt TIME} from $x$ to $x+1$. Fourth,
$(x,abs)@time$ expresses the sending of $x$ to {\tt TIME}, followed
by not sending anything to {\tt TIME} (i.e. an {}``absent'' signal).
In general, $(a,b)@time$ is syntactically correct when $a,b$ $\in$
$\mathbb{N}\cup\{ abs\}$. Fifth, $(0,1)@tick$ describes a positive
{\tt TICK}. In general, $(a,b)@tick$ is syntactically correct when
$a,b$ $\in$ $\{0,1\}$. Sixth, $(init,stop)@mode$ expresses that
the system changes from mode $init$ to mode $stop$. In general,
$(a,b)@mode$ is correct when $a,b$ $\in$ $\{ init,\, stop,\, start\}$;
its intended meaning will become clear later.

\subsection{\label{sub:Logic-of-Change-of-Change}Logic of Change of Change and
Beyond}

The Stopwatch is a prime example of modal behaviour: pressing a button
of the Stopwatch can have a different effect, depending on the mode
of operation. While the other two case studies in this paper only
contain one mode of operation, the Stopwatch contains several: $cha_{init}$,
$cha_{stop}$, $cha_{start}$, $cha_{basic}^{2}$, $cha_{reset}^{4}$,
$\ldots$ We present some intuition about these modes, before defining
the Logic of Change of Change (i.e. $\ChaChaLo$) and its extensions.
The meaning of each mode will become apparent in Section~\ref{sub:Some-Specifications-Stopwatch}.

The modes $cha_{init}$, $cha_{stop}$, and $cha_{start}$ are expressed
as simple $\ChaLo$ formulae. Mode $cha_{basic}^{2}$, on the other
hand, is expressed in $\ChaChaLo$, describing transformations between
modes $cha_{init}$, $cha_{stop}$, and $cha_{start}$. That is, $cha_{basic}^{2}$
describes an hierarchical mode, containing the simpler modes $cha_{init}$,
$cha_{stop}$, and $cha_{start}$, as is illustrated graphically in
Figure~\ref{fig:completeStopwatch}(ii). Formula $cha_{reset}^{4}$
is expressed in $\ChaFourLo$ and describes transformations between
$\ChaChaLo$ formulae. That is, $cha_{reset}^{4}$ describes an hierarchical
mode, containing simpler modes (e.g. $cha_{basic}^{2}$), as is graphically
illustrated in Figure~\ref{fig:completeStopwatch}(iii).  This hierarchical
extrapolation continues with $\ChaEightLo$ formula $cha_{lap}^{8}$,
describing transformations between $\ChaFourLo$ formulae. In general,
we deal with a $\ChaNLo$ formula with $\N$ $=$ $2^{k}$ where $k$
is strictly positive. 

A $\ChaChaLo$ formula $U$ is semantically interpreted as a pair
of changes: \\
\begin{tabular}{ll}
$s,\,\left(ch_{in},ch_{out}\right)\:\models\: U$ with:&
$s$ $\in$ $Asgmt$ ~~~and~~~$f\! ree(U)$ $\subseteq$ $(dom\: s)$~~~and\tabularnewline
&
$ch_{in},ch_{out}$ $\in$ $Change\::=\: S\! nshot\times S\! nshot$\tabularnewline
\end{tabular}\\
The pair $\left(ch_{in},ch_{out}\right)$, called a \emph{transformation},
denotes the change of $ch_{in}$ into $ch_{out}$. The definition
of $\ChaChaLo$ in Table~\ref{cap:SyntaxSemantics2Lower} is self
explanatory; we stress the similarity with $\ChaLo$ in Table~\ref{cap:SyntaxSemantics2Upper}.

\begin{table*}

\caption{\label{cap:SyntaxSemantics2Lower}$\ChaChaLo$}

\begin{center}{\scriptsize }\begin{tabular}{cc}
\hline 
{\scriptsize }\begin{tabular}{clllclcccc|ccc}
{\scriptsize (1)}&
{\scriptsize $U,\: V$}&
{\scriptsize $::=$}&
\multicolumn{10}{l}{{\scriptsize $ExprRel\:\mid\:\left(P,\: Q\right)\:\mid\: f\! alse\:\mid\: U;V\:\mid$}}\tabularnewline
{\scriptsize (2)}&
&
&
\multicolumn{10}{l}{{\scriptsize $U\Rightarrow V\:\mid\: U\,\ast\, V\:\mid\: U\,\otimes\, V\:\mid\:\exists x.\, U$}}\tabularnewline
{\scriptsize (3)}&
\multicolumn{12}{l}{{\scriptsize where $P$ and $Q$ are $\ChaLo$ formulae.}}\tabularnewline
\multicolumn{13}{l}{{\scriptsize Sugar:}}\tabularnewline
{\scriptsize (4)}&
\multicolumn{3}{r}{{\scriptsize $\left(\left(a,b\right),\left(c,d\right)\right)@p$}}&
{\scriptsize $\equiv$}&
{\scriptsize $\left(\,\left(a,b\right)@p,\left(c,d\right)@p\,\right)$}\tabularnewline
{\scriptsize (5)}&
\multicolumn{3}{r}{{\scriptsize $\neg U$}}&
{\scriptsize $\equiv$}&
{\scriptsize $U\Rightarrow f\! alse$}\tabularnewline
&
\multicolumn{3}{r}{{\scriptsize $\ldots$}}&
{\scriptsize $\ldots$ }&
{\scriptsize $\ldots$}\tabularnewline
\multicolumn{13}{l}{{\scriptsize Notation}\textbf{\scriptsize :}}\tabularnewline
\multicolumn{13}{l}{{\scriptsize }\begin{tabular}{llrccccccc|ccc}
{\scriptsize (6)}&
{\scriptsize If}&
{\scriptsize $ch$}&
{\scriptsize $=$}&
\multicolumn{9}{l}{{\scriptsize $\left(shot_{in},shot_{out}\right)$}}\tabularnewline
{\scriptsize (7)}&
{\scriptsize then}&
{\scriptsize $\left(f\! st\: ch\right)$}&
{\scriptsize $=$}&
\multicolumn{9}{l}{{\scriptsize $shot_{in}$}}\tabularnewline
{\scriptsize (8)}&
{\scriptsize and}&
{\scriptsize $\left(snd\: ch\right)$}&
{\scriptsize $=$}&
\multicolumn{9}{l}{{\scriptsize $shot_{out}$}}\tabularnewline
\end{tabular}}\tabularnewline
\multicolumn{13}{l}{\textbf{\scriptsize Semantics:}}\tabularnewline
\multicolumn{13}{l}{{\scriptsize }\begin{tabular}{lllcccc}
{\scriptsize (9)}&
\multicolumn{6}{l}{{\scriptsize $s,\,\left(ch_{in},ch_{out}\right)\:\models\: Expr\! Rel$}}\tabularnewline
&
&
&
{\scriptsize iff}&
&
&
{\scriptsize $\left(\left\llbracket Expr\! Rel\right\rrbracket \; s\right)$}\tabularnewline
\end{tabular}}\tabularnewline
\multicolumn{13}{l}{{\scriptsize }\begin{tabular}{lllcccc}
{\scriptsize (10)}&
\multicolumn{6}{l}{{\scriptsize $s,\,\left(ch_{in},ch_{out}\right)\:\models\:\left(P,\, Q\right)$}}\tabularnewline
&
&
&
{\scriptsize iff}&
&
&
\tabularnewline
&
\multicolumn{4}{l}{ {\scriptsize $s,\, ch_{in}\:\models\: P$ }}&
\multicolumn{2}{c}{{\scriptsize and}}\tabularnewline
&
\multicolumn{4}{l}{{\scriptsize $s,\, ch_{out}\:\models\: Q$}}&
\multicolumn{2}{c}{{\scriptsize and}}\tabularnewline
&
\multicolumn{6}{l}{{\scriptsize $\left(snd\: ch_{in}\right)=\left(f\! st\: ch_{out}\right)$}}\tabularnewline
\end{tabular}}\tabularnewline
\multicolumn{13}{l}{{\scriptsize }\begin{tabular}{lllcccc|}
{\scriptsize (11)}&
{\scriptsize $s,\,\left(ch_{in},ch_{out}\right)\:\models\: f\! alse$}&
&
&
\multicolumn{3}{c}{{\scriptsize never}}\tabularnewline
\end{tabular}}\tabularnewline
\multicolumn{13}{l}{{\scriptsize }\begin{tabular}{llllcccc}
{\scriptsize (12)}&
\multicolumn{7}{l}{{\scriptsize $s,\,\left(ch_{in},ch_{out}\right)\:\models\: U;V$}}\tabularnewline
&
&
&
{\scriptsize iff}&
&
&
&
\tabularnewline
&
\multicolumn{6}{l}{{\scriptsize $\exists ch_{tmp}\in Change.$}}&
\tabularnewline
&
\multicolumn{6}{l}{ {\scriptsize $s,\,\left(ch_{in},ch_{tmp}\right)\:\models\: U$ }}&
{\scriptsize and}\tabularnewline
&
\multicolumn{6}{l}{ {\scriptsize $s,\,\left(ch_{tmp},ch_{out}\right)\:\models\: V$ }}&
\tabularnewline
\end{tabular}}\tabularnewline
\end{tabular}&
{\scriptsize }\begin{tabular}{llllrclllc}
\multicolumn{10}{l}{{\scriptsize }\begin{tabular}{lllcccc}
{\scriptsize (13)}&
\multicolumn{6}{l}{{\scriptsize $s,\,\left(ch_{in},ch_{out}\right)\:\models\: U\Rightarrow V$}}\tabularnewline
&
&
{\scriptsize iff}&
&
&
&
\tabularnewline
&
{\scriptsize if }&
\multicolumn{5}{l}{{\scriptsize $s,\,\left(ch_{in},ch_{out}\right)\:\models\: U$}}\tabularnewline
&
{\scriptsize then }&
\multicolumn{5}{l}{{\scriptsize $s,\,\left(ch_{in},ch_{out}\right)\:\models\: V$}}\tabularnewline
\end{tabular}}\tabularnewline
\multicolumn{10}{l}{{\scriptsize }\begin{tabular}{lllcccc}
{\scriptsize (14)}&
\multicolumn{6}{l}{{\scriptsize $s,\,\left(ch_{in},ch_{out}\right)\:\models\: U\,\ast\, V$}}\tabularnewline
&
&
&
{\scriptsize iff}&
&
&
\tabularnewline
&
\multicolumn{6}{l}{{\scriptsize $\exists ch_{in}^{1},\, ch_{in}^{2}\in Change.$}}\tabularnewline
&
&
\multicolumn{4}{l}{{\scriptsize $ch_{in}^{1}\:\sharp\: ch_{in}^{2}$}}&
{\scriptsize and}\tabularnewline
&
&
\multicolumn{4}{l}{{\scriptsize $ch_{in}=ch_{in}^{1}\:\centerdot\: ch_{in}^{2}$}}&
{\scriptsize and}\tabularnewline
&
\multicolumn{6}{l}{{\scriptsize $\exists ch_{out}^{1},\, ch_{out}^{2}\in Change.$}}\tabularnewline
&
&
\multicolumn{4}{l}{{\scriptsize $ch_{out}^{1}\:\sharp\: ch_{out}^{2}$}}&
{\scriptsize and}\tabularnewline
&
&
\multicolumn{4}{l}{{\scriptsize $ch_{out}=ch_{out}^{1}\:\centerdot\: ch_{out}^{2}$}}&
{\scriptsize and}\tabularnewline
&
\multicolumn{5}{l}{{\scriptsize $s,\,\left(ch_{in}^{1},\, ch_{out}^{1}\right)\:\models\: U$}}&
{\scriptsize and}\tabularnewline
&
\multicolumn{5}{l}{{\scriptsize $s,\,\left(ch_{in}^{2},\, ch_{out}^{2}\right)\:\models\: V$}}&
\tabularnewline
\end{tabular}}\tabularnewline
\multicolumn{10}{l}{{\scriptsize }\begin{tabular}{lllcccc}
{\scriptsize (15)}&
\multicolumn{6}{l}{{\scriptsize $s,\,\left(ch_{in},ch_{out}\right)\:\models\: U\,\otimes\, V$}}\tabularnewline
&
&
&
{\scriptsize iff}&
&
&
\tabularnewline
&
\multicolumn{6}{l}{{\scriptsize $\exists ch_{in}^{0},\, ch_{in}^{1},\, ch_{in}^{2}\in Change.$}}\tabularnewline
&
&
\multicolumn{4}{l}{{\scriptsize $\qquad ch_{in}=ch_{in}^{0}\centerdot ch_{in}^{1}\centerdot ch_{in}^{2}$}}&
{\scriptsize and}\tabularnewline
&
\multicolumn{6}{l}{{\scriptsize $\exists ch_{out}^{0},\, ch_{out}^{1},\, ch_{out}^{2}\in Change.$}}\tabularnewline
&
&
\multicolumn{4}{l}{{\scriptsize $\qquad ch_{out}=ch_{out}^{0}\centerdot ch_{out}^{1}\centerdot ch_{out}^{2}$}}&
{\scriptsize and}\tabularnewline
&
\multicolumn{5}{l}{{\scriptsize $s,\,\left(ch_{in}^{0}\centerdot ch_{in}^{1},\, ch_{out}^{0}\centerdot ch_{out}^{1}\right)\models U$}}&
{\scriptsize and}\tabularnewline
&
\multicolumn{5}{l}{{\scriptsize $s,\,\left(ch_{in}^{0}\centerdot ch_{in}^{2},\, ch_{out}^{0}\centerdot ch_{out}^{2}\right)\models V$}}&
\tabularnewline
\end{tabular}}\tabularnewline
\multicolumn{10}{l}{{\scriptsize }\begin{tabular}{lllcccc}
{\scriptsize (16)}&
\multicolumn{6}{l}{{\scriptsize $s,\,\left(ch_{in},ch_{out}\right)\:\models\:\exists x.\, U$}}\tabularnewline
&
&
&
{\scriptsize iff}&
&
&
\tabularnewline
&
\multicolumn{6}{l}{{\scriptsize $\exists v\in V\! al.\;\; s\left[x\mapsto v\right],\,\left(ch_{in},ch_{out}\right)\:\models\: U$}}\tabularnewline
\end{tabular}}\tabularnewline
\end{tabular}\tabularnewline
\hline
\end{tabular}\end{center}
\end{table*}

Continuing in the same manner, a $\ChaFourLo$ formula, such as $\left(U,\, V\right)$,
is semantically interpreted as a pair of a pair of changes:\\
\begin{tabular}{llllllll}
\multicolumn{7}{l}{$s,\,\left(\left(ch{}_{1},ch_{2}\right),\left(ch{}_{3},ch{}_{4}\right)\right)\:\models\:\left(U,\, V\right)$}&
\tabularnewline
&
&
iff&
&
&
&
&
\tabularnewline
\multicolumn{4}{l}{ $s,\,\left(ch{}_{1},ch{}_{2}\right)\:\models\: U$ }&
and&
$s,\,\left(ch{}_{3},ch{}_{4}\right)\:\models\: V$&
and&
$\left(snd\: ch_{2}\right)=\left(f\! st\: ch_{3}\right)$\tabularnewline
\end{tabular}{\footnotesize }\\
where $U$ and $V$ are $\ChaChaLo$ formulae. $\ChaFourLo$'s complete
definition is obvious and omitted from this paper. The same remark
holds for $\ChaEightLo$ (a pair of a pair of a pair of changes) or,
in general, $\ChaNLo$ with $\N$ $=$ $2^{k}$ where $k$ is strictly
positive. The logics $\ChaChaLo$ and $\ChaFourLo$ are used in Section~\ref{sub:Some-Specifications-Stopwatch}
to capture the preemption mechanisms of the Stopwatch.

\subsubsection*{Conventions}

We present two conventions. First, an underscore denotes a \emph{don't
care} value. E.g., $(\_,0)@counter$ abbreviates $\exists x\,(x,0)@counter$.
Likewise, $(\_,\_)@counter$ abbreviates $\exists x\,\exists y\,(x,y)@counter$.

Second, similar to Section~\ref{sub:Some-Illustrations-of-ChaLo-Subway},
the implication $\left(0,1\right)@strp$ $\rightarrow$ $\left(0,0\right)@reset$
abbreviates: \\
$\left(0,1\right)@strp\:*\:\left(\_,\_\right)@reset$ ~~$\Rightarrow$~~
$\left(\_,\_\right)@strp\:*\:\left(0,0\right)@reset$. 

The previous remark holds for any of the logics. Consider for instance
$\ChaChaLo$ and the following expression:\\
$\left(\,\left(0,1\right),\left(\_,\_\right)\,\right)@strp$ $\:\rightarrow\:$
$\left(cha_{stop},cha_{start}\right)\vee\left(cha_{start},cha_{stop}\right)$
\\
and suppose $cha_{stop}$ and $cha_{start}$ only describe changes
of {\tt TICK}, {\tt COUNTER}, and {\tt TIME}. Then this expression
is an abbreviation for:\\
$\left(\,\left(0,1\right),\left(\_,\_\right)\,\right)@strp\,\ast\,\left(\,\left(\_,\_\right),\left(\_,\_\right)\,\right)@tick\,\ast\,$
\\
$\left(\,\left(\_,\_\right),\left(\_,\_\right)\,\right)@counter\,\ast\,\left(\,\left(\_,\_\right),\left(\_,\_\right)\,\right)@time$\\
~~$\Rightarrow$~~ $\left(\,\left(\_,\_\right),\left(\_,\_\right)\,\right)@strp\,*\,\left[\,\left(cha_{stop},cha_{start}\right)\vee\left(cha_{start},cha_{stop}\right)\,\right]$

\subsection{\label{sub:Some-Specifications-Stopwatch}Some Specifications}

We start by specifying the behavior of a Basic Stopwatch, which is
similar to the Stopwatch in Figure~\ref{fig:completeStopwatch}(i)
except that the {\tt RESET} button is excluded. The Basic Stopwatch's
behavior is visualized by the $\ChaChaLo$ diagram in Figure~\ref{fig:completeStopwatch}(ii).
The diagram distinguishes between three modes of operation: $cha_{init}$,
$cha_{stop}$, and $cha_{start}$. After an initialization phase,
corresponding to $cha_{init}$, the system enters a loop, executing
either mode $cha_{stop}$ or $cha_{start}$, depending on the user's
input. That is, by pressing {\tt STRP}, the Basic Stopwatch transitions
from mode $cha_{stop}$ to $cha_{start}$ or vice versa. This is expressed
in Figure~\ref{fig:completeStopwatch}(ii) by the label $\left(0,1\right)@strp$.
On the other hand, if {\tt STRP} is not pressed, the Basic Stopwatch
stays in its current mode (i.e. $cha_{stop}$ or $cha_{start}$). 

The three modes are clarified as follows. First, mode $cha_{init}$
amounts to setting {\tt COUNTER} to the value $0$. That  is:\\
$cha_{init}$ $:=$ $\left(init,\_\right)@mode\,*\,\left(\_,0\right)@counter$
\\
Note also that $mode$ book-keeps the current mode, which in this
case is $init$. Second, \\
(1)~~~~$cha_{stop}$ $:=$ $cha_{stop}^{emit}\,;\, cha_{stop}^{await}$.
\\
The first change $cha_{stop}^{emit}$ expresses that the value of
the {\tt COUNTER} stays the same and it's value $x$ has to be emitted
to {\tt TIME}: \\
$cha_{stop}^{emit}$ $:=$ $\left(stop,\_\right)@mode\,*\,\exists x.\,\left[\,\left(x,x\right)@counter\,*\,\left(\_,x\right)@time\,\right]$\\
Since the value $x$ only has to be emitted once, $cha_{stop}^{emit}$
is immediately followed in (1) by $cha_{stop}^{await}$, which expresses
that an absent signal $abs$ is sent to {\tt TIME}:\\
$cha_{stop}^{await}$ $:=$ $\left(stop,\_\right)@mode\,*\,\exists x.\:\left[\,\left(x,x\right)@counter\,*\,\left(\_,abs\right)@time\,\right]$\\
Third, \\
(2)~~~~$cha_{start}$ $:=$ $cha_{start}^{1}\wedge cha_{start}^{2}$.
\\
The first conjunct $cha_{start}^{1}$ expresses that, at every positive
{\tt TICK}, the value of {\tt COUNTER} is incremented by one (from
$x-1$ to $x$) and sent to {\tt TIME}:\\
\begin{tabular}{lll}
$cha_{start}^{1}$&
$:=$&
$\left(start,\_\right)@mode\,*\,\left(0,1\right)@tick\,*\,$\tabularnewline
&
&
$\exists x.\:\left[\,\left(x\!-\!1,x\right)@counter\,*\,\left(\_,x\right)@time\,\right]$\tabularnewline
\end{tabular}\\
The second conjunct in (2) states that, in the absence of a positive
{\tt TICK}, the value of {\tt COUNTER} remains constant and an absent
signal is sent as output:\\
\begin{tabular}{lll}
$cha_{start}^{2}$&
$:=$&
$\left(start,\_\right)@mode\,*\,\left[\left(0,0\right)@tick\,\vee\,\left(1,\_\right)@tick\right]\,*\,$\tabularnewline
&
&
$\exists x.\,\left(x,x\right)@counter\,*\,\left(\_,abs\right)@time$\tabularnewline
\end{tabular}

Having defined the $\ChaLo$ formulae, we now define the $\ChaChaLo$
formulae of Figure~\ref{fig:completeStopwatch}(ii) in three steps.
First, transformation $trans\! f_{1}$ expresses the unconditional
transition from $cha_{init}$ to $cha_{stop}$: \\
\begin{tabular}{lll}
$trans\! f_{1}$&
$:=$&
$\left(cha_{init}\,*\,(\_,\_)@time,\: cha_{stop}\right)\,*\,$\tabularnewline
&
&
$\left(\,\left(\_,\_\right),\left(\_,\_\right)\,\right)@strp\,*\,\left(\,\left(\_,\_\right),\left(\_,\_\right)\,\right)@tick$\tabularnewline
\end{tabular}\\
That is, after the initialization phase (i.e. $cha_{init}$) has taken
place, we automatically end up in $cha_{stop}$. Second, $trans\! f_{2}$
expresses that when pressing button {\tt STRP}, a transition can
take place from $cha_{stop}$ to $cha_{start}$ or vice versa:\\
$trans\! f_{2}$ $:=$ $t_{A}$ $\wedge$ $t_{B}$ \\
with:\\
 {\small }\begin{tabular}{lllllll}
{\small $t_{A}$}&
{\small $:=$}&
\multicolumn{3}{l}{{\small $\left(\,\left(stop,\_\right),\left(\_,\_\right)\,\right)@mode$ }}&
{\small $\;*\;$}&
{\small $\left(\,\left(0,1\right),\left(\_,\_\right)\,\right)@strp$}\tabularnewline
&
&
&
{\small $\rightarrow$}&
\multicolumn{3}{l}{{\small $\left(cha_{stop}*\left(\_,\_\right)@tick,\, cha_{start}\right)$}}\tabularnewline
{\small $t_{B}$}&
{\small $:=$}&
\multicolumn{3}{l}{{\small $\left(\,\left(start,\_\right),\left(\_,\_\right)\,\right)@mode$ }}&
{\small $\;*\;$}&
{\small $\left(\,\left(0,1\right),\left(\_,\_\right)\,\right)@strp$}\tabularnewline
&
&
&
{\small $\rightarrow$}&
\multicolumn{3}{l}{{\small $\left(cha_{start},\, cha_{stop}*\left(\_,\_\right)@tick\right)$}}\tabularnewline
\end{tabular}\\
Third, $trans\! f_{3}$ expresses that when button {\tt STRP} is
not pressed, the current mode stays the same:\\
$trans\! f_{3}$ $:=$ $t_{C}$ $\wedge$ $t_{D}$\\
with:\\
{\small }\begin{tabular}{lllllll}
{\small $t_{C}$}&
{\small $:=$}&
\multicolumn{3}{l}{{\small $\left(\,\left(stop,\_\right),\left(\_,\_\right)\,\right)@mode$ }}&
{\small $\;*\;$}&
{\small $\sim\left(\,\left(0,1\right),\left(\_,\_\right)\,\right)@strp$}\tabularnewline
&
&
&
{\small $\rightarrow$}&
\multicolumn{3}{l}{{\small $\left(cha_{stop}*\left(\_,\_\right)@tick,\, cha_{stop}*\left(\_,\_\right)@tick\right)$}}\tabularnewline
{\small $t_{D}$}&
{\small $:=$}&
\multicolumn{3}{l}{{\small $\left(\,\left(start,\_\right),\left(\_,\_\right)\,\right)@mode$ }}&
{\small $\;*\;$}&
{\small $\sim\left(\,\left(0,1\right),\left(\_,\_\right)\,\right)@strp$}\tabularnewline
&
&
&
{\small $\rightarrow$}&
\multicolumn{3}{l}{{\small $\left(cha_{start},\, cha_{start}\right)$}}\tabularnewline
\end{tabular}\\
and where $\sim\left(\,\left(0,1\right),\left(\_,\_\right)\,\right)@strp$
abbreviates:\\
{\small $\left(\,\left(0,0\right),\left(\_,\_\right)\,\right)@strp$
$\vee$ $\left(\,\left(1,0\right),\left(\_,\_\right)\,\right)@strp$
$\vee$ $\left(\,\left(1,1\right),\left(\_,\_\right)\,\right)@strp$.}{\small \par}

Finally, the complete behavior of the Basic Stopwatch is formalized
by:\\
$cha_{basic}^{2}$ $:=$ $trans\! f_{1}\,;\,\left(trans\! f_{2}\,\wedge\, trans\! f_{3}\right)$

\subsubsection*{Basic Stopwatch with Reset}

We now enhance the behavior of the Basic Stopwatch by including the
{\tt RESET} button. Every time {\tt RESET} is pressed, the Stopwatch
re-initializes and starts executing from the beginning, i.e. from
$cha_{init}$. This modal behaviour is illustrated by the $\ChaFourLo$
diagram in Figure~\ref{fig:completeStopwatch}(iii) where the dotted
box is a copy of Figure~\ref{fig:completeStopwatch}(ii), depicting
the hierarchical mode $cha_{basic}^{2}$. The preemptive transitions,
outside the box, have higher priority than the transitions inside
the box. That is, pressing {\tt RESET} has higher priority than pressing
{\tt STRP}. Every time {\tt RESET} is pressed, the mode $cha_{init}$
is re-executed. Formally:\\
\begin{tabular}{lll}
$cha_{reset}^{4}$&
$:=$&
$\left(\,\left(\,\left(\_,\_\right),\left(0,1\right)\,\right),\left(\,\left(\_,\_\right),\left(\_,\_\right)\,\right)\,\right)@reset$
$\:\rightarrow$\tabularnewline
&
&
$\left(cha_{basic}^{2},cha_{basic}^{2}\right)$ $\otimes$ $\left(\,\left(\,\left(\_,\_\right),\left(\_,init\right)\,\right),\left(\,\left(\_,\_\right),\left(\_,\_\right)\,\right)\,\right)@mode$\tabularnewline
\end{tabular}

\subsubsection*{Findings}

To conclude the Stopwatch case study, note that Separation Logic was
not originally intended to express the modal behavior of a system,
such as that of the Stopwatch. The above specifications seem to suggest,
however, that Separation Logic may come in handy in at least two unexpected
ways. First, the presented textual specifications denote the meaning
of the graphical diagrams in Figure~\ref{fig:completeStopwatch}(ii)
and (iii). These diagrams can be made (i.e. specified) by means of
a graphical user interface. The corresponding graphical-specification
process, in turn, could be a complementary (or competitive) alternative
for the textual-based $\Esterel$ specification process.  Second,
since the presented logics elegantly capture modal behavior, they
can of course also be used to provide an alternative formal semantics
of languages such as $\Esterel$~\cite{Berry-Gonthier-ScCompProg92}.

\section{\label{sec:Switch}Switch}

Our third case study is a $2\!\times\!2$ Switch. Its shared-memory
behavior was originally%
\footnote{If the reader is unfamiliar with $\Bluespec$, he or she can also
think of $\TLAplus$~\cite{LamportBook} as an alternative specification
language for the $2\!\times\!2$ Switch.%
} specified in the guarded-command language $\Bluespec$~\cite{BSwhitepaperSwitch}.
In this section, however, we introduce the Logic of Hierarchical Snapshots
and reuse $\ChaLo$ (cf. Table~\ref{cap:SyntaxSemantics2Upper})
to specify the Switch's behavior. 

This section consists of three parts. First, we present the design
intent of the Switch in Section~\ref{sub:Design-Intent-Switch}.
Second, we introduce the Logic of Hierarchical Snapshots in Section~\ref{sub:Logic-of-Hierarchical-Snapshots}.
Finally, we partially specify the Switch's behavior in Section~\ref{sub:Some-Specifications-Switch}.

\subsection{\label{sub:Design-Intent-Switch}Design Intent}

\begin{figure}
\begin{center}\includegraphics[%
  scale=0.5]{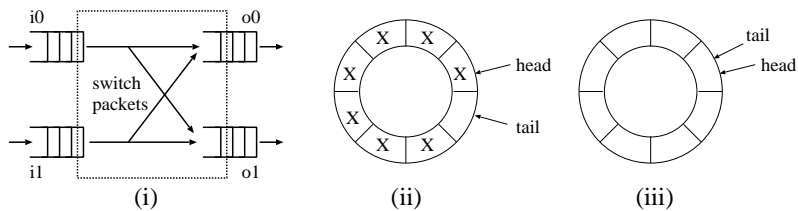}\end{center}

\caption{\label{fig:switch1}(i)~The $2\!\times\!2$ Switch, (ii)~a full
$\FIFO$ buffer, and (iii)~an empty $\FIFO$ buffer.}
\end{figure}

The $2\!\times\!2$ Switch in Figure~\ref{fig:switch1}(i) contains
two input $\FIFO$s ($\iOne$ and $\iTwo$) and two output $\FIFO$s
($\oOne$ and $\oTwo$). A data packet can arrive on $\iOne$ or $\iTwo$.
If the first bit of that packet has the value $0$, then it is routed
to $\oOne$, else to $\oTwo$. Each $\FIFO$ has the capacity to store
$1021$ data packets and $3$ management packets (see below). Each
packet contains $32$ bits. A data packet can only move if the output
$\FIFO$ is not full.  A shared resource collision can occur when
the data packets at the head of both input $\FIFO$s have the same
destination buffer (i.e. shared memory). In this case, $\iOne$ is
given priority and $\iTwo$'s data packet is delayed.

The three management packets (of each $\FIFO$) are the $head$ and
$tail$ pointers and the empty entry in Figure~\ref{fig:switch1}(ii).
The $head$ pointer refers to the entry in the $\FIFO$ that contains
the head data packet (if any). The $tail$ pointer refers to the first
empty entry. To distinguish a full $\FIFO$ from an empty $\FIFO$
(cf. Figure~\ref{fig:switch1}(iii)), one buffer entry is not used
to store a data packet. This entry, hence, stores the third management
packet of the $\FIFO$. We also mention that the $head$ and $tail$
pointers are stored in buffer entries $1022$ and $1023$, respectively.

\subsection{\label{sub:Logic-of-Hierarchical-Snapshots}Logic of Hierarchical
Snapshots}

The Logic of Snapshots has the purpose to concisely describe hierarchical
storage. We present examples below, omitting the obvious but lengthy
formal definitions. 

Suppose input buffer $\iOne$ is assigned to photographer $\PhOne$.
Then $\PhOne$ can zoom in on, say, entry number $3$ of $\iOne$
and take a snapshot of the stored packet \emph{pack}. If \emph{pack}
resembles the number five, then $\PhOne$ observes $5@i0.3$. $\PhOne$
can zoom in further by taking a snapshot of, say, the first two bits
of \emph{pack}. $\PhOne$ would then observe: $1@i0.3.0$ $*$ $0@i0.3.1$.
The first conjunct expresses that the very first bit (index $0$)
has the value one. The second conjunct states that the second bit
(index $1$) has the value zero. This indeed corresponds to the bit
notation of the number $5$, which is $0\ldots0101$ with the least
significant bit (index $0$) being the rightmost bit. If $\PhOne$
chooses to observe $d=10$ consecutive bits of \emph{pack}, starting
from bit index $n=1$, then we may write $2@i0.3.1\!-\!10$ because
these ten bits, $0\ldots010$, resemble the number two. The general
notation is: $v@i0.3.n\!-\! n'$ with $n':=n+d-1$, and where $v$
is the corresponding value. Finally, note that $\PhOne$ can also
combine disjoint snapshots of $\iOne$ as for example: \\
$1@i0.3.0$ $*$ $0@i0.3.1$ $*$ $29@i0.8$ $*$ $9@i0.12.4\!-\!17$.

\begin{figure}
\begin{center}\includegraphics[%
  scale=0.4]{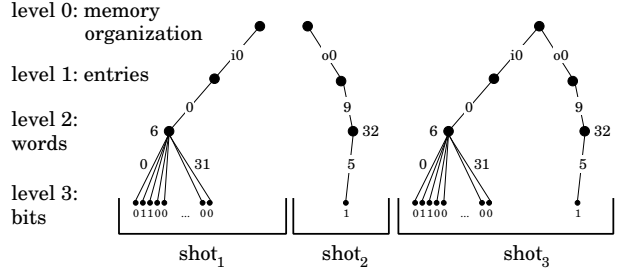}\end{center}

\caption{\label{fig:buffertree2}Three semantic snapshots.}
\end{figure}

\subsubsection*{The Semantics of Hierarchical Snapshots}

$S\! nshot$, the domain of semantic snapshots for the $2\!\times\!2$
Switch, is defined in terms of $Tree$, a parameterized semantic algebra:
\begin{tabular}{rcl}
$shot$&
$\in$&
$S\! nshot\,:=\, Tree\left[4,1024,32\right]$\tabularnewline
\end{tabular}{\footnotesize }\\
The first parameter refers to the $4$ buffers in Figure~\ref{fig:switch1}(i).
Each buffer contains $1024$ entries of $32$ bits each. 

Instead of giving a lengthy definition of $Tree$, we illustrate three
semantic snapshots in Figure~\ref{fig:buffertree2}. For example,
$shot_{2}$ is the semantic snapshot that models the syntactic snapshot
$1@o0.9.5$.  

Some more concepts follow. A path is a concatenation of edge numbers,
such as $o0.9.5$. The trace of a tree $shot$ is the set of paths
that characterize all the level~3 nodes (i.e. bits) of $shot$. E.g.:\\
\begin{tabular}{rclll}
$T\! race\left(shot_{1}\right)$&
$=$&
\multicolumn{3}{l}{$\left\{ i0.0.0,\: i0.0.1,\:\ldots,i0.0.31\right\} $}\tabularnewline
$T\! race\left(shot_{2}\right)$&
$=$&
\multicolumn{3}{l}{$\left\{ o0.9.5\right\} $}\tabularnewline
\end{tabular}{\footnotesize }\\
Two trees are disjoint ($\sharp$) iff their traces are disjoint:\\
\begin{tabular}{rclll}
$shot_{a}\:\sharp\: shot_{b}$&
$\:\textrm{iff}\:$&
\multicolumn{3}{l}{$T\! race(shot_{a})\cap T\! race(shot_{b})=\emptyset$}\tabularnewline
\end{tabular}{\footnotesize }\\
Thus, $shot_{1}\:\sharp\: shot_{2}$ holds. Since $shot_{1}$ and
$shot_{2}$ are disjoint, they can be combined ($\centerdot$) into
$shot_{3}$ as follows:\\
\begin{tabular}{rclll}
$shot_{3}$&
$=$&
$shot_{1}\centerdot shot_{2}$&
\multicolumn{2}{l}{~~with}\tabularnewline
$T\! race\left(shot_{3}\right)$&
$=$&
\multicolumn{3}{l}{$T\! race\left(shot_{1}\right)\cup T\! race\left(shot_{2}\right)$}\tabularnewline
\end{tabular}{\footnotesize }\\
Indeed, $shot_{3}$ in Figure~\ref{fig:buffertree2} represents the
combination of $shot_{1}$ and $shot_{2}$. It captures the contents
of entry number $0$ of buffer $\iOne$ and bit number $5$ of entry
number $9$ of buffer $\oOne$. Finally, when two non disjoint trees
are combined, then $\bot$ is returned. E.g.: $shot_{1}$ $\centerdot$
$shot_{1}$ $=$ $\bot$.

\subsection{\label{sub:Some-Specifications-Switch}Some Specifications}

In conformance to the hierarchical snapshots, presented in the previous
section, we now present $\ChaLo$ specifications of the Switch. 

As a first example,  we want $idleI\! nputBu\! f\!\left[bu\! f\right]$
to state that no packet is taken out of input buffer $bu\! f$ --where
$bu\! f$ is $i0$ or $i1$. In accordance to Figure~\ref{fig:switch1}(ii-iii),
we therefore want to specify that $bu\! f$'s head pointer $head$
does not change. Since $head$ is stored in $bu\! f\!.\! f\! ir\! st$,
we write the following where $bu\! f$ is $i0$ or $i1$: \\
~(i)~~$idleI\! nputBu\! f\!\left[bu\! f\right]$ $\equiv$ $\exists head.\;\left(head,head\right)@bu\! f\!.\! f\! ir\! st$

As a second example, we want $retrieveF\! romBu\! f\!\left[bu\! f\right][n_{1}][n_{2}]\left[value\right]$,
with $bu\! f$ equal to $i0$ or $i1$, to state that: value $value$
corresponds to $value@bu\! f\!.head.n_{1}-n_{2}$ where $head$ is
the head pointer of $bu\! f$. That is, we want to retrieve (but not
extract) the $n_{2}-n_{1}+1$ bits, starting from index $n_{1}$,
from the head data packet in $bu\! f$. Formally, we have:\\
\begin{tabular}{cllllllll}
(ii)&
\multicolumn{8}{l}{$retrieveF\! romBu\! f\!\left[bu\! f\right][n_{1}][n_{2}]\left[value\right]$}\tabularnewline
(iii)&
&
$\equiv$&
$\exists\, head.$&
$(\;$&
\multicolumn{3}{l}{$\left(head,\_\right)@bu\! f\!.\! f\! ir\! st$}&
\tabularnewline
(iv)&
&
&
&
&
&
$\ast$&
$\left(value,\_\right)@bu\! f\!.head.n_{1}\!-\! n_{2}$&
$)$\tabularnewline
\end{tabular}\\
Note that (iii) does not specify the new contents of $bu\! f\!.\! f\! ir\! st$
and (iv) does not specify the new contents of $bu\! f\!.head.n_{1}\!-\! n_{2}$. 

Based on (ii), we can now define the following:\\
~(v)~~$retrieveF\! romBu\! f\!\left[bu\! f\right]\left[value\right]$
$\equiv$ $retrieveF\! romBu\! f\!\left[bu\! f\right][0][31]\left[value\right]$
\\
where $bu\! f$ is $i0$ or $i1$. That is, $value$ represents the
complete data packet that is stored at the head of $bu\! f$.

As a third example, we want $extractF\! romBu\! f\!\left[bu\! f\right][n_{1}][n_{2}]\left[value\right]$,
with $bu\! f$ equal to $i0$ or $i1$, to be similar to $retrieveF\! romBu\! f\!\left[bu\! f\right][n_{1}][n_{2}]\left[value\right]$,
except that we now not only retrieve but also extract the data bits
from the  head packet in $bu\! f$. Formally:\\
\begin{tabular}{cllllllll}
(vi)&
\multicolumn{8}{l}{$extractF\! romBu\! f\!\left[bu\! f\right][n_{1}][n_{2}]\left[value\right]$}\tabularnewline
(vii)&
&
$\equiv$&
$\exists\, head.\,$&
$(\:$&
\multicolumn{3}{l}{$\left(head,\left(1+head\right)mod\,1022\right)@bu\! f\!.\! f\! irst$}&
\tabularnewline
(viii)&
&
&
&
&
&
$\ast$&
$\left(value,\_\right)@bu\! f\!.head.n_{1}\!-\! n_{2}$&
$)$\tabularnewline
\end{tabular}\\
Note that in (vii) we now do specify the new contents of $bu\! f\!.\! f\! ir\! st$.

Based on (vi), we can  define the following where $bu\! f$ is $i0$
or $i1$:\\
~(ix)~~$extractF\! romBu\! f\!\left[bu\! f\right]\left[value\right]$
$\equiv$ $extractF\! romBu\! f\!\left[bu\! f\right][0][31]\left[value\right]$

An additional remark is that, constraints, such as: \\
\begin{tabular}{llccccccc|ccl}
\multicolumn{11}{l}{(x)}&
$\exists x.\: extractF\! romBu\! f\!\left[bu\! f\right]\left[x\right]\,\rightarrow\, notEmptyBu\! f\!\left[bu\! f\right]$\tabularnewline
\end{tabular}{\footnotesize }\\
also have to be specified. In words, (x) states that extracting a
packet $x$ from buffer $bu\! f$ implies that $bu\! f$ is not empty.
The trivial definition of $notEmptyBu\! f$ is omitted. 

As a fourth example, consider:\\
\begin{tabular}{lllccccccc|ccc|}
(xi)&
$\left(depart:x,0\right)@i0$&
$\equiv$&
\multicolumn{10}{l}{$extractF\! romBu\! f\!\left[i0\right][0][31]\left[x\right]$}\tabularnewline
&
&
&
$\otimes$&
\multicolumn{9}{l}{$extractF\! romBu\! f\!\left[i0\right][0][0][0]$}\tabularnewline
\end{tabular}\\
It states that $\iOne$'s head packet $x$ is extracted from the buffer
and that it's first bit has the value $0$. Similarly:\\
\begin{tabular}{llllccccccc|ccc|}
(xii)&
$\left(arrive:y,0\right)@o0$&
\multicolumn{2}{l}{$\equiv$}&
\multicolumn{10}{l}{$insertI\! nBu\! f\!\left[o0\right][0][31]\left[y\right]$}\tabularnewline
&
&
\multicolumn{2}{l}{}&
$\otimes$&
\multicolumn{9}{l}{$insertI\! nBu\! f\!\left[o0\right][0][0]\left[0\right]$}\tabularnewline
\end{tabular}{\footnotesize }\\
The definition of $insertI\! nBu\! f$ is omitted from this paper.

Based on the above, we now define:\\
\begin{tabular}{lllccccc|ccc|}
(xiii)&
\multicolumn{10}{l}{$\exists z.\,\left(\:\left(depart:z,0\right)@i0\,\rightarrow\,\left(arrive:z,0\right)@o0\:\right)$}\tabularnewline
\end{tabular}{\footnotesize }\\
This expresses, amongst other things, that the departed packet~$x$
and the arrived packet~$y$ are one and the same packet~$z$. Finally,
consider:\\
\begin{tabular}{lllccccc|ccc|}
(xiv)&
\multicolumn{10}{l}{$\exists z.\,\left(\:\left(arrive:z,0\right)@o0\,\rightarrow\,\left(depart:z,0\right)@i0\,\vee\,\left(depart:z,0\right)@i1\:\right)$}\tabularnewline
\end{tabular}\\
The arrival of a packet at $\oOne$ implies its departure from $\iOne$
or $\iTwo$. Continuing in this manner, we can completely capture
the Switch's behavior.

\subsubsection*{Findings}

To conclude the $2\!\times\!2$ Switch case study, note that Separation
Logic is typically used to verify pointer-intensive code~\cite{O'Hearn,Reynolds}.
Since the Switch also contains pointers, it is less surprising, compared
to the Stopwatch case study, that Separation Logic can be used as
a specification language for shared-memory systems such as the Switch.

\section{\label{sec:Conclusions}Conclusions \& Future Work }

We have captured the concurrent behavior of three very different systems
by means of Separation Logic and its natural extensions. Instead of
specifying a modal-based system in $\Esterel$ and a shared-memory
system in $\Bluespec$, we are now able to specify both systems by
means of the same formalism --not to mention the Subway system which
was originally specified in $\LUSTRE$. That is, we have a unifying
framework for multiple design approaches that initially seemed disparate.
Alternatively, we could (in future work) provide a semantics for $\LUSTRE$,
$\Esterel$, and $\Bluespec$ in our unifying formalism.

Critics may remark that any other specification language, say $\Esterel$,
can also be used to capture the behavior of any of the three presented
systems. Hence, they might question the relevance of the formalism,
presented in this paper. We respond in the two following ways.

First, we have provided insight into how three seemingly independent
concurrent systems are related to each other: (i)~the Switch's behavior
merely differs from the Subway's behavior in that it requires hierarchical
snapshots instead of plain snapshots, and (ii)~the Stopwatch's behavior
merely differs from the Subway's in that it requires change of change
(and change of change of change) to be specified instead of only change.
Now, in our formalism, anything that can be expressed with plain snapshots
can also be expressed with hierarchical snapshots. Similarly, anything
that can be expressed with change (cf. $\ChaLo$) can also be expressed
with the more powerful concept of change of change (cf. $\ChaChaLo$),
etc. So, all three concurrent systems, presented in this paper, can
be expressed in one and the same formalism which we denote here (for
the first time) by: $\ChaSpeLo$, which is an instantiation of $\ChaNHLo$.
The parameters $\N$ and $\h$ denote the number of changes and the
hierarchical depth, respectively. For example, $n=h=1$ for the Subway,
$n=4$ and $h=1$ for the Basic Stopwatch with Reset, and $n=1$ and
$h=3$ for the Switch. 

\begin{quote}
The potential power of our formalism $\ChaNHLo$ lies in being able
to select a specific subset, defined by the values of $\N$ and $\h$,
for a given application domain. \emph{}
\end{quote}
Second, we invite the reader to check whether the other specification
languages (e.g. $\Esterel$) can in fact capture the behavior of all
three concurrent systems in a uniform and sufficiently concise way.
As mentioned in the introduction, practitioners will typically not
use $\Esterel$ to specify a shared-memory system and will not use
$\Bluespec$ to specify the modal behavior of e.g. a Stopwatch. 

Finally, in line with this paper, we also refer to our complementary
work~\cite{DaylightFM09} in which we have applied different specification
languages (including $\Bluespec$) to one and the same case study
(i.e. the $2\!\times\!2$ Switch case study).

\subsubsection*{Acknowledgement}

This work was partially supported by an AFOSR grant and NSF grant
CCF-0702316.


\begin{thebibliography}{10}
\bibitem{Walker1}{\small A. Ahmed, L. Jia, D. Walker, {}``Reasoning about Hierarchical
Storage'',} \emph{\small LICS'03}{\small . }{\small \par}
\bibitem{BSwhitepaperSwitch}{\small \char`\"{}Automatic Generation of Control Logic with Bluespec
SystemVerilog\char`\"{}, Feb. 2005 Bluespec, Inc., Available at:}
\emph{\small bluespec.com/products/Papers.htm }{\small \par}
\bibitem{Berry-Gonthier-ScCompProg92}{\small G. Berry, G. Gonthier,} \emph{\small {}``The $\Esterel$
Synchronous Programming Language: Design, Semantics, Implementation''}{\small ,
Science of Computer Programming, 1992.  }{\small \par}
\bibitem{CairesCardelli}{\small L. Caires, L. Cardelli, {}``A Spatial Logic for Concurrency'',}
\emph{\small TACS 2001}{\small , pp. 1-37.}{\small \par}
\bibitem{DaylightFM09}{\small E.G. Daylight, S.K. Shukla, {}``On the Difficulties of Concurrent-System
Design, Illustrated with a 2x2 Switch Case Study'', accepted for}
\emph{\small Formal Methods 2009}{\small , Eindhoven, the Netherlands,
November 2-6, 2009.}{\small \par}
\bibitem{FokkinkBook}{\small W.J. Fokkink,} \emph{\small {}``Introduction to Process Algebra''}{\small ,
Texts in Theoretical Computer Science, Springer, 2000.}{\small \par}
\bibitem{HalbwachsBook}{\small N. Halbwachs,} \emph{\small {}``Synchronous Programming of
Reactive Systems''}{\small , Kluwer'93.}{\small \par}
\bibitem{HalbwachsSubway}{\small N. Halbwachs, F. Lagnier, C. Ratel,} \emph{\small ``Programming
and verifying real-time systems by means of the synchronous data-flow
language $\LUSTRE$''}{\small , IEEE Transactions on Software Engineering'92. }{\small \par}
\bibitem{Statecharts}{\small D. Harel,} \emph{\small \char`\"{}Statecharts: A Visual Formalism
for Complex Systems\char`\"{}}{\small , Science of Computer Programming,
1987, 8, pp. 231-274.  }{\small \par}
\bibitem{HoareComSeqProc}{\small C.A.R. Hoare,} \emph{\small {}``Comm. Seq. Processes''}{\small ,
Comm. ACM 21, 1978, pp. 666-677. }{\small \par}
\bibitem{LamportBook}{\small L. Lamport,} \emph{\small {}``Specifying Systems: The $\TLAplus$
Language and Tools for Hardware and Software Engineers'',} {\small Addison-Wesley
Professional, 2002. }{\small \par}
\bibitem{MilnerPicalculus}{\small R. Milner,} \emph{\small {}``Communicating and Mobile Systems:
the $\pi$-calculus''}{\small , Cambridge University Press, 1999.
}{\small \par}
\bibitem{O'Hearn}{\small P.W. O'Hearn, J.C. Reynolds, H. Yang, {}``Local Reasoning
about Programs that Alter Data Structures'',} \emph{\small CSL}{\small '01.}{\small \par}
\bibitem{SynchHypothesis}{\small D. Potop-Butucaru, R. de Simone, J-P. Talpin, {}``The Synchronous
Hypothesis and Synchronous Languages'', in R. Zurawski, ed.,} \emph{\small {}``The
Embedded Systems Handbook}{\small '', CRC Press, 2005.}{\small \par}
\bibitem{Reynolds}{\small J.C. Reynolds, {}``Separation Logic: A Logic for Shared Mutable
Data Structures.'',} \emph{\small LICS}{\small '02. }{\small \par}
\bibitem{Yang07RelationalSepLog}{\small H. Yang,} \emph{\small {}``Relational Separation Logic''}{\small ,
Theoretical Computer Science, Vol. 375, Issue 1-3, May 2007, pp. 308-334.
 }
\end{thebibliography}
\end{document}